\documentclass[aps,pre,twocolumn,showpacs,amsmath,amsmath,amssymb,amsfonts]{revtex4-1}
\usepackage{graphicx}
\usepackage{dcolumn}
\usepackage{bm}
\usepackage{color}
\usepackage{multirow}
\usepackage{mathtools}
\DeclarePairedDelimiter\bra{\langle}{\rvert}
\DeclarePairedDelimiter\ket{\lvert}{\rangle}
\DeclarePairedDelimiterX\braket[2]{\langle}{\rangle}{#1 \delimsize\vert #2}

\begin{document}
\title{Self-similar non-equilibrium dynamics of a many-body system\\ with power-law interactions}

\author{Ricardo Guti\'errez}
\author{Juan P. Garrahan}
\author{Igor Lesanovsky}
\affiliation{School of Physics and Astronomy, University of
Nottingham, Nottingham, NG7 2RD, UK}

\date{\today}
\begin{abstract}
The influence of power-law interactions on the dynamics of many-body systems far from equilibrium is much less explored than their effect on static and thermodynamic properties. To gain insight into this problem we introduce and analyze here an out-of-equilibrium deposition process in which the deposition rate of a given particle depends as a power-law on the distance to previously deposited particles. This model draws its relevance from recent experimental progress in the domain of cold atomic gases which are studied in a setting where atoms that are excited to high-lying Rydberg states interact through power-law potentials that translate into power-law excitation rates. The out-of-equilibrium dynamics of this system turns out to be surprisingly rich. It features a self-similar evolution which leads to a characteristic power-law time dependence of observables such as the particle concentration, and results in a scale invariance of the structure factor. Our findings show that in dissipative Rydberg gases out of equilibrium the characteristic distance among excitations --- often referred to as the blockade radius --- is not a static but rather a dynamic quantity.

\end{abstract}


\maketitle

\section{Introduction}

Self-similar behavior is ubiquitous in physics. It emerges for example in condensed matter systems near criticality \cite{Chaikin2000,*Sachdev2007}, in biological systems \cite{Brown2000} as well as in complex networks \cite{Leland1994,*Park2000,*Song2005}. Systems with this property ``look the same'' under a rescaling of characteristic length and/or time scales. This property often entails a drastic reduction of the details that are necessary for a theoretical description, such as specific parameter values or initial conditions. Self-similar behavior can also occur in cold atomic ensembles, currently used as an experimental platform for exploring fundamental questions of far-from-equilibrium physics such as the approach of interacting many-body systems to (thermal) equilibrium \cite{Kinoshita2006,Trotzky2012,Langen2015}. For example, self-similar behavior is predicted for the relaxation of a gas of quenched Bose-condensed atoms as a consequence of the existence of non-thermal fixed points \cite{Berges2008,Orioli2015}.  Moreover, it is argued \cite{Schmiedmayer2013} that many of these characteristics were also present during the evolution of the early universe leading to the idea of using cold atomic systems as analogue systems for addressing problems of relevance to cosmology or particle physics.

In this work we introduce a simple far-from-equilibrium scenario in which a non-trivial relaxation dynamics is driven by power-law interactions. The motivation stems from recent experiments in the domain of cold atom physics which explore the laser excitation of atoms to high-lying electronic states in which they interact with power-law potentials. While complementary, the scenario we study below has analogies with that of the above-mentioned work on Bose-condensed gases \cite{Berges2008,Schmiedmayer2013,Orioli2015}. In particular, evolution which is initially fast and uncorrelated is succeeded by slow and strongly correlated self-similar growth. We show that the physics of this dynamics is governed by an effective particle deposition process where the rates depend on the distance to other deposited particles as a power-law. The scale-invariant nature of the deposition dynamics is revealed analytically and confirmed numerically by means of extensive numerical simulations. Moreover, we show that the relaxation dynamics crosses over into a mean-field regime when the power-law exponent becomes equal to the system dimension. Beyond providing an understanding of non-equilibrium processes governed by power-law interactions our results reveal new insights into the relaxation behavior of gases of interacting Rydberg atoms, which are currently widely employed for the experimental study of many-body phenomena \cite{Schauss2012,Carr2013,Schempp2014,Malossi2014,Barredo2015,Weber2015,Urvoy2015}. Our results indicate that the characteristic minimal distance between Rydberg excitations --- which is often referred to as blockade radius --- is not generally a static quantity but can in the presence of dissipation acquire a non-trivial scale-invariant time-dependence. 

\begin{figure*}[t]
\begin{minipage}[b]{0.32\textwidth}
\includegraphics[scale=0.26]{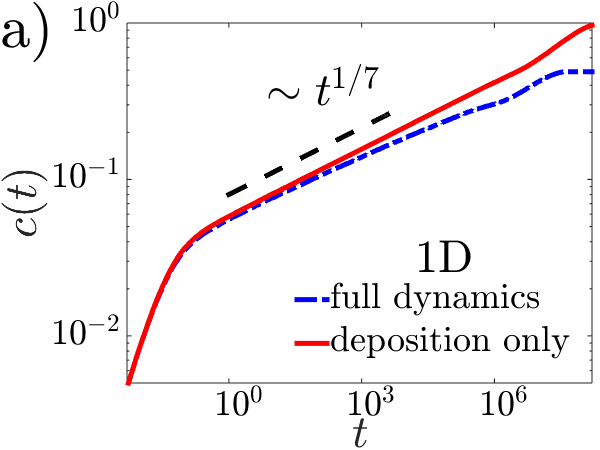}
\end{minipage}
\begin{minipage}[b]{0.32\textwidth}
\includegraphics[scale=0.26]{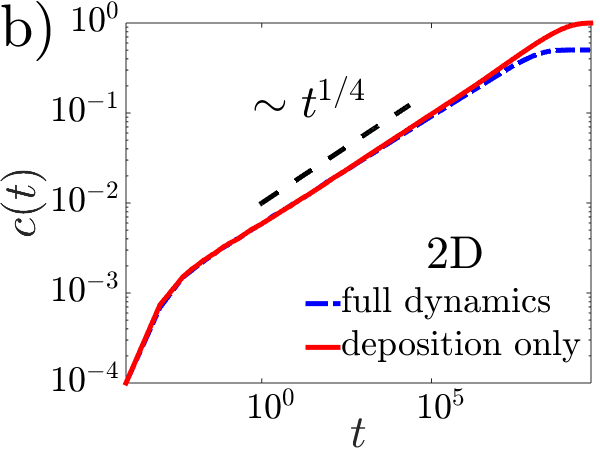}
\end{minipage}
\begin{minipage}[b]{0.32\textwidth}
\includegraphics[scale=0.21]{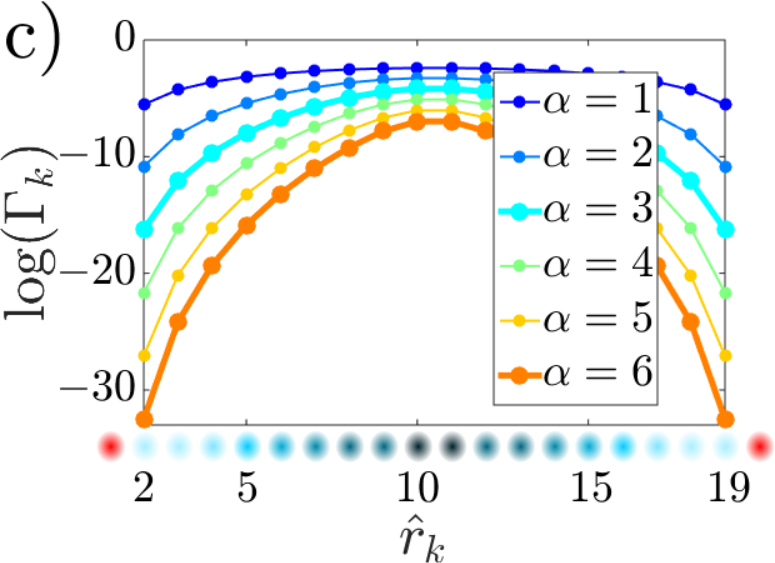}
\end{minipage}\\
\vspace{-0.2cm}
\begin{center}
\includegraphics[scale=0.7]{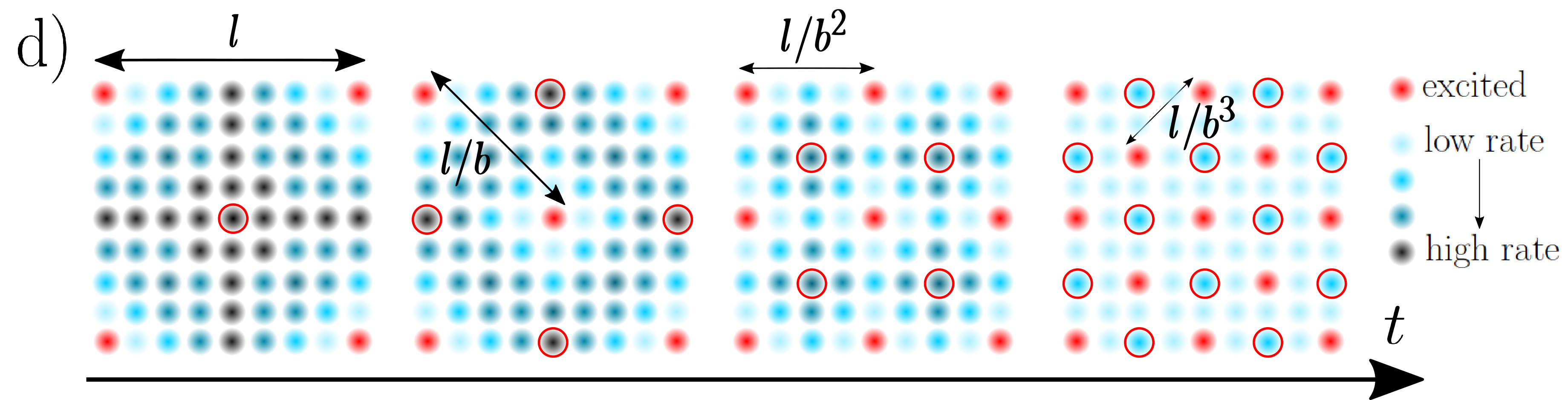}
\end{center}
\vspace{-0.4cm}
\caption{ (Color online) {\sf \bf Deposition process vs. full Rydberg gas dynamics.}
(a) Concentration $c(t)$ of excitations as a function of time in a 1D chain of $N=10^4$ atoms interacting through a power-law potential ($\alpha = 3$, $R=15$) for the full process (deposition and removal; blue dashed line) and for pure deposition (red continuous line). The black dashed segment corresponds to a power-law with exponent $d/(2\alpha + d) = 1/7$. See text for details. (b) Same results for a $N=100\times 100$ 2D square lattice. The black dashed segment corresponds to a power-law of exponent $1/4$. (c) Natural logaritm of the deposition rate $\Gamma_k$ [Eq. (\ref{rates})] as a function of $\hat{r}_k$ in a 1D lattice ($R=15$) with excitations being present on sites $1$ and $20$. (The most relevant cases, $\alpha = 3$ and $6$, are accentuated, and the order in which the lines appear starting from the top, $\alpha=1$, and moving down to the bottom line, $\alpha=6$, is that of increasing $\alpha$ values.) (d) Sketch of the deposition model in a 2D lattice. Red (blue) discs represent excited (ground state) atoms. The darker the blue the higher the deposition rate for subsequent excitations. Encircled discs indicate the most likely positions for deposition.}\label{fig1}
\end{figure*}

\section{Effective dynamics}

The dynamical problem we study below emerges from considering a collection of $N$ atoms on a cubic lattice in $d$ dimensions. Each atom is described with two levels, a ground state and the excited Rydberg state, denoted as $\ket{\downarrow}$ and $\ket{\uparrow}$ respectively. Atoms in the Rydberg state at positions ${\bf r}_k$ and ${\bf r}_m$ interact through a power-law potential $V^{(\alpha)}_{km} = C_\alpha/|{\bf r}_k - {\bf r}_m|^\alpha$ with exponent $\alpha$, while the much weaker interactions involving atoms in the ground state are neglected. Typically encountered exponents are $\alpha=6$ (van der Waals interaction) and $\alpha=3$ (dipolar interaction) \cite{Saffman2010}. The atomic states $\ket{\uparrow}$ and $\ket{\downarrow}$ are resonantly coupled by a laser field and at the same time subject to strong dissipation which leads to the rapid dephasing of superposition states, e.g. $\ket{\uparrow}+\ket{\downarrow}$. The evolution of the system is governed by a (classical) spin-flip dynamics in which a change in the state of site $k$, whether an excitation $\ket{\downarrow} \to \ket{\uparrow}$ or a de-excitation $\ket{\uparrow} \to \ket{\downarrow}$, occurs with an (operator-valued) rate (see Ref. \cite{Lesanovsky2013})
\begin{equation}
\Gamma_k^{-1} = 1 + \left[R^\alpha \sum_{m\neq k} \frac{n_m}{|\hat{{\bf r}}_k - \hat{{\bf r}}_m|^\alpha}\right]^2.
\label{rates}
\end{equation}
Here $R$ \footnote{The relation between $R$ and the parameters in the microscopic description of the system in term of the quantum master equation is given in \cite{Lesanovsky2013}.} parameterizes the interaction strength, $n_k=\left|\uparrow\right>_k\!\left<\uparrow\right|$ is the excitation number operator and $\hat{{\bf r}}_k = {\bf r}_k/a$ are position vectors in units of the lattice constant $a$. The term $R^\alpha \sum_{m\neq k} n_m/|\hat{{\bf r}}_k - \hat{{\bf r}}_m|^\alpha$ strongly correlates the atoms, giving rise to a slowing down of the dynamics in the vicinity (within a distance of the order of $R$) of an excited atom. In this sense, the dynamics is affected by kinetic constraints of the type that are usually considered in simplified models of the glass transition \cite{Ritort2003,*Chandler2010}. This description of the Rydberg gas dynamics has been recently used in a number of theoretical works \cite{Garttner2012,Petrosyan2013,Petrosyan2013b,Lesanovsky2013,Lesanovsky2014,Marcuzzi2014,Hoening2014} and was also successfully employed to model experiments \cite{Schempp2014,Urvoy2015}. A similar perturbative approach to the derivation of effective dynamics in the limit of strong dissipation was proposed in Ref. \cite{Cai2013}.

We consider an \lq empty\rq\, initial state, $\ket{\mathcal{C}} = \ket{\downarrow \downarrow \downarrow \cdots \downarrow}$. One key observation that greatly helps to simplify the following analysis is that excitation processes $\ket{\downarrow} \to \ket{\uparrow}$ strongly dominate over de-excitations $\ket{\uparrow} \to \ket{\downarrow}$, as will be shown below. Thus, the effective reversible dynamics is in essence a pure (and irreversible) {\it deposition process} (where $\ket{\uparrow} \to \ket{\downarrow}$ is not allowed) governed by the following master equation
\begin{equation}
\partial_t \ket{P(t)} = \sum_k \Gamma_k \left[\sigma_+^k - (1 - n_k) \right] \ket{P(t)}.
\label{deposition_eq}
\end{equation}
The state is given by the vector $\ket{P(t)} \equiv \sum_\mathcal{C} P(\mathcal{C};t) \ket{\mathcal{C}}$, with $P(\mathcal{C};t)$ denoting the probability of finding a specific configuration of excitations $\ket{\mathcal{C}}$ at time $t$. The operator $\sigma_+^k$ creates an excitation at site $k$, or --- in the deposition picture --- deposits a particle in site $k$. Figure \ref{fig1} shows the concentration of excitations $c(t) \equiv \sum_k \langle n_k(t) \rangle / N$ as a function of time for the case of $\alpha=3$. We see from Figs.\ \ref{fig1}(a) and \ref{fig1}(b) [1D chain and a 2D square lattice, respectively] that irreversible deposition processes indeed dominate the growth regime, similar to what was suggested recently in Ref. \cite{Sanders2015}. Deviations become apparent in the long-time limit when the system approaches its stationary state: for the reversible process this is the fully random state, $\lim_{t\to\infty} c(t) = 1/2$, while for the deposition process it is the (absorbing) state of a full lattice, $\lim_{t\to\infty} c(t) = 1$. Note, that in the case of Rydberg gases this long-time limit will typically not be achieved due to the finite lifetime of excited atoms. However, while a finite decay rate eventually makes the system settle into a nonequilibrium stationary state, the initial phases of the growth dynamics are well within the reach of current experiments, and they turn out to be virtually indistinguishable from those observed without decay, according to both numerical \cite{Lesanovsky2013} and experimental evidence \cite{Valado2015}.

\section{Deposition model}

\begin{figure}[ht]
\hspace{-0.8cm}\includegraphics[scale=0.22]{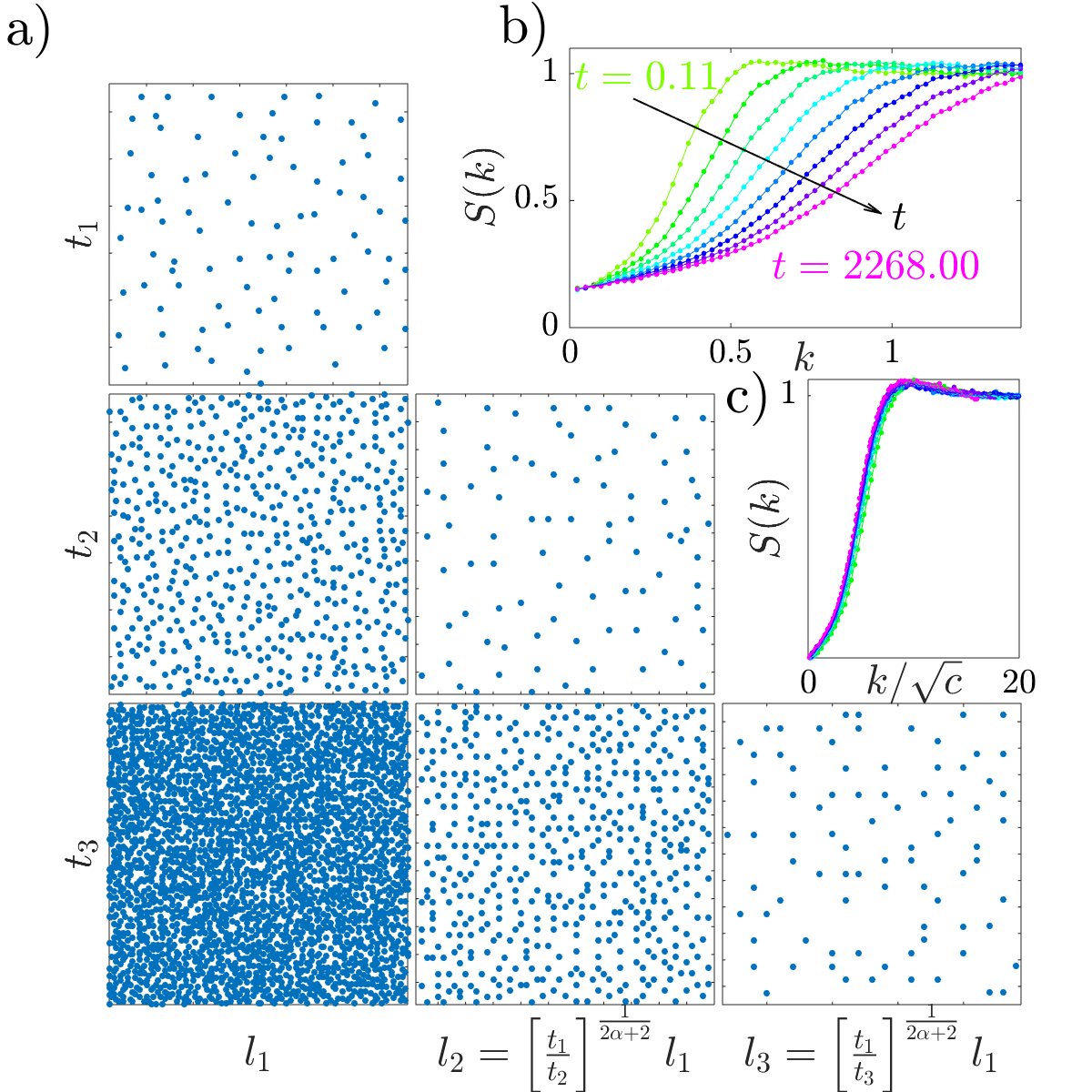}
\vspace{-0.2cm}
\caption{(Color online) {\sf \bf Scale invariant evolution.} (a) Excitations (blue dots) in a 2D square lattice of $N=256\times 256$ atoms, $\alpha = 3$ and $R=15$. The first column shows a subsystem of size $l_1 \times l_1$ sites for $l_1 = 128$ at three different times: $t_1 = 1$, $t_2 = 10^3$, $t_3 = 10^6$. The second column shows the configuration within a box of linear size $l_2 = (t_1/t_2)^{1/(2\alpha + 2)} l_1 \simeq 54$ at times $t_2$ and $t_3$. The third column shows the configuration within a box of linear size $l_3 = (t_1/t_3)^{1/(2\alpha + 2)} \simeq 23$ at time $t_3$. (b)  Snapshots of the structure factor $S(k)$ taken at different times. (c)  $S(k)$ as a function of the scaled wavenumber $k/\sqrt{c(t)}$. All the lines shown in panel (b) collapse onto a single curve.}\label{fig2}
\end{figure}

In the strongly correlated regime, i.e. when the interparticle distance becomes smaller than $R$, the deposition process is amenable to an approximate analytical treatment. This is most transparently explained in 1D: Consider two excitations in a 1D chain separated by distance $l$. Then the next excitation will be deposited with very high likelihood at (essentially) a distance $l/2$ from the two existing excitations, effectively rescaling the inter-excitation distance by the factor $b = 2$. The reason for this is that the deposition rate is highly peaked at the mid point between two excitations as shown in Fig. \ref{fig1}(c). The process continues until the natural cutoff scale --- the lattice spacing $a$ --- is reached, which makes further subdivision impossible. A similar process is expected to occur in 2D, where the same reasoning, starting for example from a square of excitations, can be applied with a rescaling constant $b = \sqrt{2}$ [see Fig. \ref{fig1}(d)]. That the actual dynamics indeed follows a rescaling behavior is nicely observed in Fig. \ref{fig2}(a) (details further below).

For a more quantitative analysis we consider the distribution $\pi(l,t)$ of distances $l$ between nearest excitations. Formalizing the above considerations suggests that the dynamics of the deposition process is approximately described by the following master equation
\begin{equation}
\partial_t \pi(l,t) = \frac{(bl)^d}{N} \Gamma(bl) \pi(bl,t) - \frac{l^d}{N} \Gamma(l) \pi(l,t),
\label{Smastereq} 
\end{equation}
where $\pi(l,t)$ is the probability distribution of the distance between nearest excitations $l$  at time $t$. Here, $d = 1$, $2$, or $3$ gives the dimensions of space and $b = 2/\sqrt{d}$ is the scaling parameter. The rate at which the rescaling step $l \to l/b$ occurs is $\Gamma(l) \approx z^{-2} R^{-2\alpha} l^{2\alpha} \phi(b)$ as soon as there are excitations within distances shorter than $R$, according to Eq. (1). Here, $z$ is the coordination number of the lattice ($z=2d$ as we consider $d$-dimensional cubic lattices) and $\phi(b)$ is a geometric factor that depends on the space dimensionality. For instance, in 1D $\phi(b) = [\sum_{m=0}^{N/2-1} (1/b + m)^{-1}]^{-2\alpha}$, where different terms in the sum correspond to different \lq \lq shells\rq\rq\, of excitations. We can therefore rewrite Eq. (\ref{Smastereq}) as
\begin{equation}
\partial_\tau \pi(l,\tau) = \left(\frac{bl}{L}\right)^{(2\alpha+ d)} \pi(bl,\tau) -  \left(\frac{l}{L}\right)^{(2\alpha+ d)}  \pi(l,\tau)
\label{Smastereq2} 
\end{equation}
where $L = N^{1/d}$ is the linear size of the lattice. In order to simplify the equation, we have absorbed the factor $z^{-2} R^{-2\alpha} \phi(b) L^{2\alpha}$, which is fixed for a given lattice and interaction potential, into a rescaled time $\tau$. The fact that $\phi(b)$ does not play any role in the derivation of the self-similar behavior we report means that the initial condition shown in Fig. \ref{fig1} (d) is simply a conveniently simple example for the purpose of illustration. As the normalized inter-excitation distances $l/L$ follow $1 \to 1/b \to 1/b^2 \to \cdots$, we can rewrite Eq. (\ref{Smastereq2}) in terms of $\pi_n(\tau)$, the probability of having an inter-excitation distance $l= b^{-n} L$ for $n=0,1,2,\ldots$, 
\begin{equation}
\partial_\tau \pi_n(\tau) = b^{-(n-1)(2\alpha+d)} \pi_{n-1}(\tau) - b^{-n(2\alpha+d)} \pi_{n}(\tau), 
\label{Smastereq3} 
\end{equation}
where the rhs can be most economically expressed in terms of $f_n(\tau) \equiv b^{-n(2\alpha+d)} \pi_{n}(\tau)$ as $f_{n-1}(\tau) - f_{n}(\tau)$. \\

What we have in Equation (\ref{Smastereq3}) is the master equation in terms of $\pi(l,t)$ when the deposition is such that $l$ takes on a discrete set of values. Nevertheless, in general the initial condition $\pi(l,0)$ is not a delta function, but a random configuration that arises from the initial creation of independent, distant excitations. Moreover, the fact that sometimes excitations do not occur exactly at the maximum of the rates forces us to consider continuous distributions where the rescaling process with parameter $b$ takes place simultaneously at slightly different scales. To address these issues, we move on to the continuum,  $l=b^{-x} L$ for $x\geq 0$, and Taylor expand $f(x,t)$ (assuming analyticity with respect to $x$): 
$f(x-1,\tau) - f(x,\tau) = -\sum_{p=1}^\infty \frac{1}{p!} \left. \frac{\partial^p}{\partial x^{\prime p}} f(x^\prime,t)\right|_{x^\prime = x-1}$. 
In this setting, we can rewrite the master equation back in terms of $l$ using the function $\bar{\pi}(l,\tau) \equiv (l/L)^{(2\alpha+d)} \pi(l,\tau)$
\begin{equation}
\frac{\partial_\tau \bar{\pi}(l,\tau)}{\left(l/L\right)^{-(2\alpha+d)}}  = \sum_{p=1}^\infty \frac{(-1)^{p+1}}{p!} \left[\left(\log{(b)}\, l^\prime \partial_{l^\prime} \right)^p \bar{\pi}(l^\prime,\tau)\right]_{l^\prime = bl}.
\label{Smastereq4} 
\end{equation}
Plugging in the wave-like ansatz $\bar{\pi}(l,t) = F(\tau - g(l))$ (we expect the distribution to shift towards shorter and shorter distances as time goes by), we obtain 
\begin{eqnarray}
&\frac{F^{(1)}(\tau - g(l))}{(l/L)^{-(2\alpha+d)}}  = &-\log{b}\, F^{(1)}(\tau - g(bl))\, bl\, g^{(1)} (bl)\\ \nonumber
& &-\frac{\log^2{b}}{2}\, F^{(1)}(\tau - g(bl))\, bl\, g^{(1)} (bl)\\ \nonumber
& & -\frac{\log^2{b}}{2}\, F^{(2)} (\tau - g(bl))\left[bl\, g^{(1)}(bl)\right]^2\\ \nonumber
& &- \frac{\log^2{b}}{2}\, F^{(1)}(\tau - g(bl)) (bl)^2 g^{(2)} (bl)+ \cdots
\end{eqnarray}
stopping at order $p=2$. For higher orders on the rhs one obtains more terms that are (up to some constant factor, including functions of $b$) products of $F(\cdot)$ or its derivatives $F^{(m)}(\cdot)$ and $(bl)^m g^{(m)}(bl)$ raised to a certain power, where $g^{(m)}(bl)$ is the $m$-th derivative of $g(\cdot)$ evaluated at $bl$. As the $l$ dependence has to cancel on both sides for the equation to hold for all times, we are left with $g(l) = l^{-(2\alpha+d)}$, and therefore (after absorbing $L$ into the undetermined function $F(\cdot)$)
\begin{equation}
\pi(l,\tau) = l^{-(2\alpha+d)} F(\tau - l^{-(2\alpha+d)}).
\label{Swavelike}
\end{equation} 

The mean inter-excitation distance as a function of time, $\langle l (\tau) \rangle = \int dl\, l\, \pi(l,\tau)/\int dl\, \pi(l,\tau)$, is, after changing the integration variables to $y = \tau - l^{-(2\alpha+d)}$,
\begin{equation}
\langle l(\tau) \rangle = \tau^{-\frac{1}{2\alpha+d}} \frac{\int dy\, (1 - y/\tau)^{-\frac{2}{2\alpha+d}} F(y)}{\int dy\, (1 - y/\tau)^{-\frac{1}{2\alpha+d}} F(y)}.
\label{Smeanl}
\end{equation}
For any dilute initial condition, $F(\cdot)$ is non-zero for small values of $y$. On the other hand, the process spans several orders of magnitude in time (see Fig. \ref{fig1} (a, b)), so for sufficiently long times there has to be a strong cancellation in the argument of $F$, and $l^{-(2\alpha+d)}$ must grow essentially in parallel with $\tau$. This means that eventually $(1 - y/\tau) \approx 1$ and, writing back in terms of the original time units, $\langle l(t) \rangle \sim t^{-\frac{1}{2\alpha+d}}$, or, equivalently,
\begin{equation}
c(t) =\langle l(t) \rangle ^{-d} \sim t^{\frac{d}{2\alpha+d}}.
\label{Sconcent}
\end{equation}
This result is in agreement with the prediction obtained and verified in Ref. \cite{Lesanovsky2013} by a more restrictive reasoning that did not consider the time dependence of the full distribution $\pi(l,t)$. Eq. (\ref{Sconcent}) justifies the use of the exponents $1/7$ and $1/4$ in Fig. \ref{fig1} (a, b). By the same reasoning we can rewrite $\pi(l,t)$ as given in Eq. (\ref{Swavelike}) as a scale invariant function for long times:
\begin{equation}
\pi(l,t) \approx l^{-(2\alpha+d)} \bar{F}(t l^{(2\alpha+d)}),
\label{scaling}
\end{equation}
showing that the full distribution of distances to nearest excitations evolves in a self-similar way. Physically, this means that the system looks the same at any time, only the characteristic distances become shorter and shorter. The prediction of self-similarity from the approximate description in terms of $\pi(l,t)$ appears robust for the actual deposition problem, as seen from kinetic Monte-Carlo simulations of Eq. (\ref{deposition_eq}): Fig. \ref{fig2}(a) shows that the configurations of a 2D system at different times, and thus very different concentration, look statistically similar if lengths are scaled according to Eq. \eqref{scaling}.  The data in Figs. \ref{fig2} (b,c) confirm this observation through the collapse of the structure factors at different times.

\begin{figure}[ht]
\includegraphics[scale=0.24]{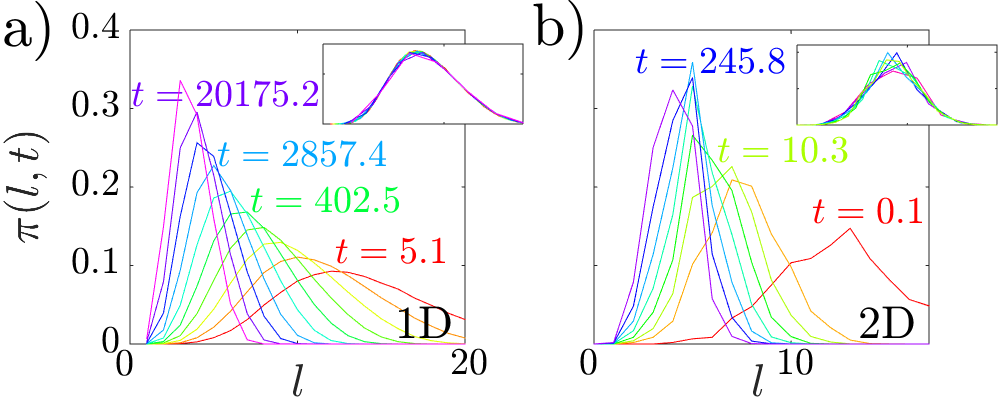}\\
\vspace{-0.2cm}
\caption{(Color online) {\sf \bf Scale invariance of $\pi(l,t)$ in 1D and 2D for $\alpha = 3$.}
(a) Distribution of distances between nearest excitations $\pi(l,t)$ and rescaling according to Eq. (\ref{scaling}) for $\alpha = 3$ in a 1D chain ($N=2^{16}$, $R=15$). Lines of different color correspond to different times. The inset shows the collapse of the distribution functions under Eq. (\ref{scaling}) [distributions are shown as a function of $l\, t^{1/(2\alpha + d)}$, and multiplied by $t^{-1/(2\alpha + d)}$ to account for the change in the measure of the distribution]. (b) Same as panel (a) for 2D ($N=256\times 256$, $R=15$).}\label{fig3}
\end{figure}

Furthermore, the precise scaling relation in Eq.\eqref{scaling} is also verified in the simulations. The numerically obtained distributions of distances between nearest excitations are shown in Fig. \ref{fig3}. The insets show the rescaled distributions, which nicely collapse onto a master curve. This works quite well even for small values of $l$, where the discreteness of the lattice introduces some roughness in the curves, especially in 2D [Fig. \ref{fig3} (b)]. 

\section{Dynamic Rydberg blockade}

\begin{figure}[ht]
\includegraphics[scale=0.26]{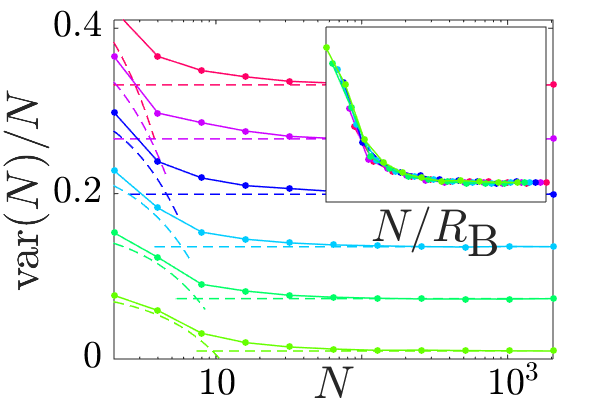}
\vspace{-0.2cm}
\caption{(Color online) {\sf \bf Excitation number variance as a function of the system size.}
Variance in a chain of $2^{16}$ atoms ($\alpha=3$, $R=15$) for (sub-)system sizes $N = 2, 4, \ldots, 2^{11}$ (averaged over 1000 realizations). The dashed curves correspond to the expression in Eq. (\ref{varmodel}) applied to each $c$, where $R_\textrm{B}$ is extracted from the largest $N$ considered. Different lines correspond to concentrations starting from approximately $c = 0.09$ (light green line at the bottom) to $c = 0.30$ (pink line on top) [lines have been displaced vertically to improve the visibility]. \emph{Inset:} Number variance curves normalized by $N c\,(1-c) \left\{1 - (2 R_\textrm{B}-1)\, c \right\}$ and plotted as functions of $N/R_\textrm{B}$.
}\label{fig4}
\end{figure}

The form of the rates \eqref{rates} leads to pronounced spatial anti-correlations, i.e. there exists a \lq\lq correlation hole\rq\rq\, between excitations, cf.\ Figs.\ \ref{fig1}(d) and \ref{fig2}(a). In the context of Rydberg gases the linear size of this correlation hole is often referred to as the blockade radius $R_\mathrm{B}$. Since the concentration increases in time $R_\mathrm{B}$ will become a dynamic quantity and thus we encounter a {\it dynamic Rydberg blockade}. The time-dependence of $R_\mathrm{B}$ can actually be inferred from a macroscopic measurement, namely the analysis of the system-size dependence of the fluctuations in the number of excitations. The excitation number variance at a given time in the non-equilibrium evolution of a system of size $N$ is $\textrm{var}(N) = \langle n(N)^2 \rangle - \langle n(N) \rangle^2$, where $n(N) = \sum_{i=1}^N n_i$. After some algebra, this can be written in terms of $\delta n_i \equiv n_i - c$ (as usual $c = \langle n_i \rangle$) as follows 
\begin{equation}
\textrm{var}(N) = N\left\{c\,(1-c) + \sum_{k\neq 0} \langle \delta n_i \delta n_{i+k}\rangle\right\}.
\label{Svariance}
\end{equation}
The anti-correlations due to the blockade effect are expected to decay at a finite distance for sufficiently large $N$. Indeed, starting from a very small $N$ (within the blockade radius), $\sum_{k\neq 0} \langle \delta n_i \delta n_{i+k}\rangle$ is expected to become more and more negative as $N$ grows, until $N$ reaches beyond the Rydberg blockade radius, $\delta n_i$ and  $\delta n_{i+k}$ for sufficiently large $|\hat{\bf r}_i - \hat{\bf r}_{i+k}|$ become independent ($\langle \delta n_i \delta n_{i+k} \rangle = 0$), and therefore $\sum_{k\neq 0} \langle \delta n_i \delta n_{i+k} \rangle$ saturates. In Fig. \ref{fig4} we show numerically obtained $\textrm{var}(N)/N$ as a function of $N$ in a 1D lattice. The fact fhat the curves become flat for sufficiently large $N$ is in agreement with the simple reasoning enunciated above. As expected, for small sizes a more complex dependence of the fluctuations on $N$ is observed. 

To understand the small $N$ behavior, we consider a hard-objects description \cite{Sun2008,Weimer2008,Ates2012,Honing2013}, i.e. we assume  that there are no two excitations within a radius $R_\textrm{B}$. For the sake of simplicity, we focus on the 1D case, although the same reasoning is expected to be relevant in lattices of higher dimensions. If $N\leq 2 R_\textrm{B}$, then $\sum_{k\neq 0} \langle \delta n_i \delta n_{i+k} \rangle = (N-1)c(1-c)(-c)$. Here, $c$ is the probability of having $n_i = 1$, i.e. an excitation at a generic site $i$, $(1-c)(-c)$ is the product of the fluctuations at $i$ ($\ket{\uparrow}$) and $i+k$ ($\ket{\downarrow}$), and  $N-1$ results from the summation. We do not consider the case $n_i = 0$ because then $\langle \delta n_i \delta n_{i+k} \rangle = -c \left[c (1-c)+(1-c)(-c)\right] = 0$, where the term in square brackets includes the fluctuation for $\ket{\uparrow}$ and $\ket{\downarrow}$ states weighted by their probability of occurrence at $i+k$ (considering a ground state atom can be located at any distance from an excitation). When $N> 2 R_\textrm{B}$, we obtain the aforementioned saturation $\sum_{k\neq 0} \langle \delta n_i \delta n_{i+k} \rangle = (2 R_\textrm{B}-1)c(1-c)(-c)$. Summarizing,
\begin{equation}
\textrm{var}(N)=\left\{
\begin{array}{cc}
N c\,(1-c) \left\{1 - (N-1)\, c \right\}, & N\leq 2 R_\textrm{B}\\
N c\,(1-c) \left\{1 - (2 R_\textrm{B}-1)\, c \right\}, & N > 2 R_\textrm{B}.
\end{array}%
\right.  \label{varmodel}
\end{equation}
Despite the simplicity of the approximation, Eq. (\ref{varmodel}) is shown to be essentially valid in Fig. \ref{fig4}, where we show $\textrm{var}(N)$ for different times together with the approximation (\ref{varmodel}). The data collapses upon rescaling the system size by $R_\mathrm{B}\propto t^{-\frac{1}{2\alpha+d}}$ highlighting its usefulness for studying the dynamical nature of the Rydberg blockade from macroscopic observations. The left hand side is calculated numerically (or it is measured in an experiment), which allows us to extract the value of $R_\mathrm{B}$.

\section{Mean-field dynamics arising from long-range interactions}

\begin{figure}[ht]
\includegraphics[scale=0.24]{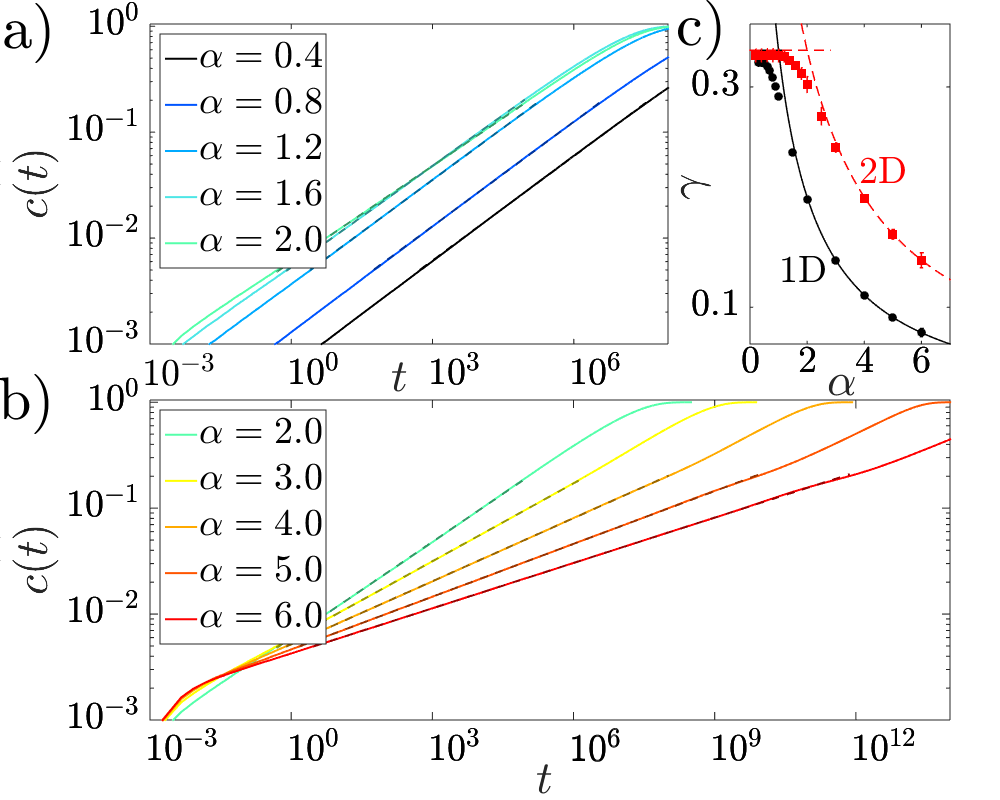}
\vspace{-0.2cm}
\caption{(Color online) {\sf \bf Time evolution of the concentration for different power-law exponents $\alpha$.}
(a) Concentration $c(t)$ in a square lattice ($N=317\times 317$, $R=15$) with long-range interactions, $\alpha \leq d=2$ (averages of 30 realizations). The larger the value of $\alpha$, the faster the dynamics (and therefore the higher the position at which the line appears in the plot). The dashed segments superimposed to each line show the result of a power-law fit $c(t) \sim t^\gamma$. (b) Same as in (a) for $\alpha \geq d$. In this case, the larger the value of $\alpha$, the slower the dynamics (and therefore the lower the position at which the line appears in the plot).  (c) Growth exponent $\gamma$ vs. interaction exponent $\alpha$ in the same 2D square lattice (red squares) and a 1D chain (black dots; $N=10^5$, $R=15$). Dashed lines correspond to $\gamma = 1/3$ (mean-field exponent, for $\alpha < d$) and $\gamma = {d/(2\alpha + d)}$ (valid for $\alpha > d$).}\label{fig5}
\end{figure}

Finally, we consider the dependence of the deposition dynamics on the power-law exponent $\alpha$. Qualitatively there are two different situations,  that of short-range interactions ($\alpha > d$) and that of long-range interactions ($\alpha \leq d$) \cite{Campa2014}. Figure \ref{fig5} shows the time evolution of the concentration in both cases.  Whenever the power-law rates are short-ranged, the growth is exponential for very short times, and then follows the scaling law $c(t) = t^{d/(2\alpha+d)}$, cf.\ Eq. \eqref{Sconcent}.  In contrast, in the long-ranged case the concentration grows in an $\alpha$-independent manner, $c(t) \sim t^{\gamma}$, with $\gamma \approx 1/3$ throughout (or, more precisely, after an extremely short exponential growth whose duration decreases with growing system size). The behavior of the concentration for $\alpha \leq d$ can indeed be understood by a mean-field analysis, as we presently aim to show.

The deposition dynamics is governed by Eq. (\ref{deposition_eq}), with rates given by Eq. (\ref{rates}). The average of the number operator corresponding to site $j$, $\langle n_j(t)\rangle = \bra{-}n_j\ket{P(t)}$, with $\ket{-} \equiv \sum_\mathcal{C} \ket{\mathcal{C}}$, evolves in time according to 
\begin{eqnarray}
&\partial_t \langle n_j(t)\rangle &= \displaystyle\sum_k  \bra{-} \Gamma_k n_j\left[\sigma_+^k - (1 - n_k) \right] \ket{P(t)} \\ \nonumber
& &=\bra{-} \Gamma_j\, n_j\left[\sigma_+^j - (1 - n_j) \right] \ket{P(t)}\\ \nonumber
& &=\bra{-}\Gamma_j\,(1-n_j) \ket{P(t)}
\label{meanfieldn}
\end{eqnarray}
where the mean-field lack of correlations between sites is exploited. Moreover, given the equivalence of all sites, $n_j(t) \to n(t)$, $\partial_t c(t) = \bra{-}\Gamma(n(t)) \,(1-n(t)) \ket{P(t)}$, where as usual $c(t) = \langle n(t)\rangle$. We can approximate this equation by replacing expectation values
of products of operators by products of expectation values
of operators: $\partial_t c(t) \approx \Gamma(c(t)) (1 - c(t))$. The resulting mean-field equation for the time evolution of the density is
\begin{equation}
\partial_t c(t) = \frac{1-c(t)}{1+[F_\alpha R^\alpha c(t)]^2}.
\label{Smf}
\end{equation}
Here, the geometric factor $F_\alpha \equiv \sum_k |\hat{\bf r}_k|^{-\alpha}$ converges for $\alpha > d$ and grows unboundedly for $\alpha \leq d$ as $N$ is increased. It can be shown that the system asymptotically reaches the stationary solution $c_s = 1$.

For $\alpha > d$, we expect the existence of an initial stage, $c(t)\ll 1$, for which the denominator in Eq. (\ref{Smf}) is essentially one, and there is an initial exponential growth 
\begin{equation}
c(t) = 1 - \exp(-t).
\label{initialgrowth}
\end{equation}
 Physically, this corresponds to the creation of independent, distant excitations that eventually become the initial seed for the deposition process.  In the case of $\alpha \leq d$, due to the unbounded growth of $F_\alpha \equiv \sum_k |\hat{\bf r}_k|^{-\alpha}$ with the system size, this regime is expected to be negligibly short for sufficiently large $N$. This is in agreement with the results reported in Fig. \ref{fig5}.

For a better understanding of the other regimes involved, we obtain the following implicit solution to Eq. (\ref{Smf}) by separation of variables: 
\begin{equation}
t = -(1+F_\alpha^2 R^{2\alpha}) \log{(1- c(t))} - \frac{F_\alpha^2 R^{2\alpha}}{2} \left(2 c(t) + c(t)^2 \right). 
\label{implicitsol_mf}
\end{equation}
For long times, $c(t)$ is approximately 1 and the logarithmic term is dominant, which gives an exponential relaxation
\begin{equation}
c(t) \approx  1 - \exp(- t/(1+F_\alpha^2 R^{2\alpha})).
\label{initialgrowth}
\end{equation}
So the final stages of the mean field dynamics are ruled by exponential growth in all cases, as seen in Fig. \ref{fig5}.

To study the intermediate time regime, for which $c(t)$ is still considerably smaller than $1$, it is useful to expand the implicit solution in powers of $c(t)$: 
\begin{equation}
t = c(t) + c(t)^2/2 +  (1 + F_\alpha^2 R^{2\alpha})  \left[c(t)^3/3+ \mathcal{O}(c^4)\right]. 
\label{implicitperturb_mf}
\end{equation}
If $[F_\alpha R^\alpha c(t)]^2 \gg 1$, we obtain the following algebraic growth:
\begin{equation}
c(t) \approx [3 /(F_\alpha^2 R^{2\alpha})]^{1/3} t^{1/3}.
\label{intermediategrowth}
\end{equation}
This is in agreement with the exponent $\gamma \approx 1/3$ observed for $\alpha \leq d$ in Fig. \ref{fig5}, as is the fact that the prefactor of the power law increases as $\alpha$ becomes larger. For long-range interactions, $\alpha \leq d$, the dynamics is indeed effectively of the mean field type. 

In conclusion, there are three distinct dynamical regimes qualitatively identical to those reported in Ref. \cite{Lesanovsky2013}. For very short times, the creation of independent, distant excitations makes the concentration grow exponentially in time (for large $N$ this is only observable if $\alpha > d$), as predicted at the mean field level. For very long times, the system reaches exponentially the stationary state. Throughout most of the non-equilibrium evolution, the concentration grows algebraically according to the self-similar dynamics, Eq. (\ref{Sconcent}), for $\alpha > d$, and following a mean-field dynamics with exponent $\gamma = 1/3$ in the long-range interacting case. 

\begin{figure}[h]
\includegraphics[scale=0.24]{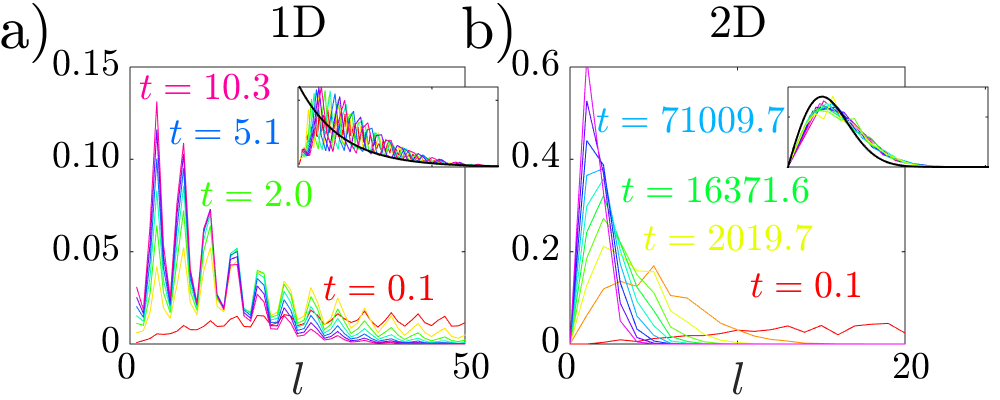}\\
\vspace{-0.2cm}
\caption{(Color online) {\sf \bf Scale invariance of $\pi(l,t)$ in 1D and 2D for $\alpha = 0.8$.}
(a) Distribution of distances between nearest excitations $\pi(l,t)$ and rescaling according to Eq. (\ref{intermediategrowth}) for $\alpha = 0.8$ in a 1D chain ($N=2^{16}$, $R=15$). Lines of different color correspond to different times. The inset shows the collapse of the distribution functions under Eq. (\ref{intermediategrowth}) [distributions are shown as a function of $l\, t^{1/3d}$, and multiplied by $t^{-1/3d}$ to account for the change in the measure of the distribution. The black line corresponds to the analytical prediction $\pi(x,t)^{(d)} = d (A/2^d) B x^{d-1} e^{-(A/2^d) B x^d}$ (see text for an explanation)]. (b) Same as panel (a) for 2D ($N=256\times 256$, $R=15$).}\label{fig6}
\end{figure}

To gain more insight into the dynamics of the mean-field regime, we study the distribution of interparticle distances $\pi(l,t)$ for $\alpha \leq d$. In the absence of spatial correlations, which is the characteristic of mean field approaches, the probability of finding zero excitations in a small spherical volume $\delta V$ (in units given by $a^d$) at time $t$ is $P(0; \delta V) = 1 - c(t) \delta V + \mathcal{O}(\delta V^2)$. Therefore, the probability of finding no excitations in a volume $V + \delta V$ is 
\begin{equation}
P(0; V + \delta V ) = P(0;V)(1 - c(t) \delta V + \mathcal{O}(\delta V^2)),
\label{Poisson}
\end{equation}
 and thus $(P(0; V + \delta V) - P(0;V))/\delta V = -c(t)$. In the limit of $\delta V\to 0$, we obtain the normalized distribution 
\begin{equation}
P(0;V)\, dV = c(t) e^{-c(t) V} dV. 
\label{Poisson}
\end{equation}
So far in this paragraph, the reasoning has followed the standard Poisson process derivation. As $V = A (l/2)^d$ (where $A$ is a geometric factor), we write in terms of $l$, and using the previously established notation, 
\begin{equation}
\pi(l,t) dl = d (A/2^d) c(t) l^{d-1} e^{-c(t) (A/2^d) l^d} dl. 
\label{Poisson_l}
\end{equation}
The time dependence can be made explicit by using the mean-field result displayed in Eq. (\ref{intermediategrowth}), $c(t) \approx B t^{1/3}$, where for convenience we define $B \equiv  [3 /(F_\alpha^2 R^{2\alpha})]^{1/3}$, resulting in 
\begin{equation}
\pi(l,t) dl = d (A/2^d) B t^{1/3} l^{d-1} e^{-(A/2^d) B t^{1/3} k l^d} dl. 
\label{Poisson_l2}
\end{equation}
In terms of a rescaled variable $x \equiv t^{1/3d} l$, the distribution can be written as
\begin{equation}
\pi(x,t)^{(d)} dx = d (A/2^d) B x^{d-1} e^{-(A/2^d) B x^d} dx,
\label{Poisson_x}
\end{equation}
where we have explicitly included the dimensions of space as a superscript in $\pi(x,t)^{(d)}$. It turns out that the full distribution is also scale invariant in time in the long-range interacting case, and a rescaling similar to that seen in the case $\alpha > d$, but with exponent $1/3d$, should thus be possible. This is indeed confirmed in Fig. \ref{fig6}, which is analogous to Fig. \ref{fig3}, but for $\alpha = 0.8$ instead of $3$. Due to the oscillations observed in 1D at small distances, we  limit ourselves to relatively early times in that case, to make the approximate adequacy of the collapse at least partly visible. These oscillations seem to indicate that the mean-field assumption does not hold perfectly, and some spatial correlations are still visible at $\alpha=0.8$ at the level of the full distribution function.

Besides knowing that there has to be an approximate scale invariance for $\alpha < d$ due to the approximate validity of Eq. (\ref{Poisson_x}), we can use that equation to calculate explicitly the functional form of the rescaled distribution. In a one dimensional chain, we obtain $\pi(x,t)^{(1)} = \pi B  e^{-\pi B x}$, while in a two-dimensional lattice the distribution is $\pi(x,t)^{(2)} = 2\pi B x  e^{-\pi B x^2}$. In these expressions, the symbol $\pi$ without explicit dependence on any variables is just the constant $\pi$ (the ratio of a circle's circumference to its diameter), and it should not be confused with the distribution $\pi(x,t)^{(d)}$.  The corresponding curves are shown as continuous black lines in the insets of  Fig. \ref{fig6} (a,b). We want to emphasize that no fitting procedure has been used, as the functional form is completely closed except for the parameter $B$, which can be obtained numerically for a lattice of a given topology and size. In fact, $B$ changes depending on the dimensionality because $F_\alpha$ does, and, for a cubic lattice of $N = 2^{16}$ sites, its value is $B=0.0198$ for $d=1$ and $B= 0.00214$ for $d=2$. While in the case of $d=1$ [Fig. \ref{fig6} (a)] the agreement is not very good (which is to be expected, as it presupposes a constant growth as $l$ becomes smaller that is not realistic in the presence of blockade effects -even in this largely, but not perfectly, mean field regime), in $d=2$ [Fig. \ref{fig6} (b)] the fit is fairly good. So it is clear that the distribution of inter-excitation distances in the long-range interacting regime also shows at least approximate scale invariance, and that the distribution is not far from that one would find in the idealized mean-field case, i.e. in the complete absence of correlations.

\section{Conclusions}

We have introduced a simple far-from-equilibrium scenario in which a non-trivial relaxation dynamics is driven by power-law interactions that result in rates that depend on the distance to existing excitations as a power-law. The motivation originates from the study of Rydberg gases far from equilibrium. We provide a simple model for the deposition process that captures the essential physics. The scale-invariance of the deposition dynamics is revealed analytically and confirmed by means of extensive numerical simulations of the original problem (without de-excitations). Our results indicate that the blockade radius acquires a scale-invariant time-dependence in the presence of dissipation. Moreover, we study how the dynamics depends on the exponent of the interactions, and show a crossover into a mean-field regime when the power-law exponent becomes equal to the dimensions of the lattice or smaller.

Despite the evident differences in the underlying physical models, it is worthy of mention that a scale-invariant dynamics has also been theoretically established in observables characterizing domain growth and phase ordering in classical models of magnetism that have been quenched from a disordered phase to an ordered phase, even in the absence of power-law interactions \cite{Derrida1995}. Intriguingly, this is another setting in which the dynamics can be thought of as a deposition process. It is not clear to us whether deeper relations between that problem and the one that concerns us here exist (though the fact that the power-law exponents differ speaks against the use of conventional universality arguments).

Whether connections to problems studied in the past are eventually found or not, the presented out-of-equilibrium setting is probably one of the simplest manifestations of a many-body evolution governed by power-law rates, and therefore we expect that a similar dynamics may be observed in systems of quite a different nature than the dissipative Rydberg gases that inspired its study. In view of the experimental realization through Rydberg atoms it is interesting to ask how much of the observed features actually persists in the quantum regime, i.e. when the strong noise condition is lifted. While computationally unfeasible this could very well be explored in the most recent generation of experiments, e.g. \cite{Schauss2012,Barredo2015,Maller2015}.

\begin{acknowledgments}
\textit{Acknowledgements.---} We would like to thank Emanuele Levi for his careful reading of the manuscript, and one anonymous reviewer for suggesting a study of the distribution of inter-excitation distances in the presence of long-range interactions, which we have included in the paper. The research leading to these results has received funding from the European Research Council under the European Union's Seventh Framework Programme (FP/2007-2013) / ERC Grant Agreement No. 335266 (ESCQUMA), the EU-FET grant HAIRS 612862 and from the University of Nottingham. Further funding was received through the H2020-FETPROACT-2014 grant No.  640378 (RYSQ). We also acknowledge financial support from EPSRC Grant no.\ EP/J009776/1. Our work has benefited from the computational resources and assistance  provided  by the University of Nottingham High Performance Computing service.
\end{acknowledgments}

\end{document}